# MODELLING THE RISE AND FALL OF TWO-SIDED MOBILITY MARKETS WITH MICROSIMULATION

**Farnoud Ghasemi, Corresponding author**
*Faculty of Mathematics and Computer Science, Jagiellonian University, Krakow, Poland*
*Email: farnoud.ghasemi@doctoral.uj.edu.pl*

**Rafal Kucharski**
*Faculty of Mathematics and Computer Science, Jagiellonian University, Krakow, Poland*
*Email: rafal.kucharski@uj.edu.pl*

## Abstract

In this paper, we propose a novel modelling framework to reproduce the market entry strategies for two-sided mobility platforms. In the MaaSSim agent-based simulator, we develop a co-evolutionary model to represent day-to-day dynamics of the two-sided mobility market with agents making rational decisions to maximize their perceived utility. Participation probability of agents depends on utility, composed of: experience, word of mouth and marketing components adjusted by agents every day with the novel S-shaped formulas – better suited (in our opinion) to reproduce market entry dynamics than previous approaches. With such a rich representation, we can realistically model a variety of market entry strategies and create significant network effects to reproduce the rise and fall of two-side mobility platforms.

To illustrate model capabilities, we simulate a 400-day evolution of 200 drivers and 2000 travelers on a road-network of Amsterdam. We design a six-stage market entry strategy with consecutive: kick-off, discount, launch, growth, maturity and greed stages. After 25 days the platform offers discounts, yet it starts gaining market share only when the marketing campaign launches at day 50. Campaign finishes after 50 days, which does not stop the growth, now fueled mainly with a positive word of mouth effect and experiences. The platform ends discounts after 200 days and reaches the steady maturity period, after which its greedy strategy leads to collapse of its market share and profit. All above simulated with a single behavioral model, which well reproduces how agents of both sides adapts to platform actions.

*Keywords*: Ride-sourcing, Two-sided platforms, Transport network companies, Market entry strategies

## 1. Introduction

Ride-sourcing companies – also known as Transportation Network Companies (TNC) – such as Uber and Didi have reached significant market share in a short time through the platform business model. For instance, in the Uber case, what began in 2009 as a car service in San Francisco is now valued at $42.55 billion and operates in more than 80 countries and 10,000 cities around the world (*1*). The reason underlying such a tremendous potential to grow in two-sided markets is the power of network. The classic definition by Rochet and Tirole (*2*) characterizes them as markets in which one or several platforms enable interactions between end-users and try to get both sides on board by appropriately charging each side. The platforms associated with these markets rely on the critical mass needed for their self-sustainable operations and the network effect to promote growth (*3*, *4*). However, these platforms apply various market entry strategies in the early adaptation phases to follow a specific growth pattern that includes different stages from launch to maturity (*5*, *6*). In terms of two-sided mobility markets, the platform must be attractive not only to travelers but also to drivers in order to gain market shares. Travelers want to be at their destinations in a short time, with high comfort, reliability, and low price. Drivers, working for the platform, want to earn enough not only to cover their costs (fuel, depreciation costs of vehicles), but to be at least on par with the so-called reservation wage (the expected minimum wage when working elsewhere). Even though two-sided mobility platforms can potentially grow rapidly, they face serious challenges to reach profitability. They must ensure minimum scale for a significant cross-side network effect that refers to the direct increase in value for users on one side of a network by adding users to another side (*7–9*). It becomes much more difficult to achieve profitability, considering that the supply and demand are decentralized (the platform does not have direct control over them). That is why platforms often opt for large subsidies and incur heavy losses at the early, adaptation phases. Trip fare, commission rate, and discounts are the main tuning levers that platform uses, in addition to marketing, to steer the supply-demand interactions toward profitability.

    Clearly, there are questions surrounding the viability of ride-sourcing companies, which calls for thorough studies on the underlying evolution and growth mechanism. The specific character of the two-sided mobility markets and their unique features led to a wide body of literature. Several scientific works examine equilibrium in the market (*10–12*), stability in the market (*13–15*), the platform pricing strategy (*16–20*), and competition between the platforms (*21*, *22*). While most of these studies are equilibrium-based, the rest consider fixed demand or supply. Hence, they cannot provide a complete image of the dynamics the platform. Some studies (*23*, *24*) develop dynamic adjustment models based on 'exponential smoothing' implemented on agent experiences. However, they do not consider the main growth factors, such as utilization through word of mouth and marketing. Besides, due to the memory-based structure, their adjustment function is not adequate enough to stabilize the agents' behavior. Empirical observations suggest perception is adjusted in a specific range and grows slowly at the extreme (in which a person is positive or negative) and fast at the neutral states. Therefore, the adjustment follows an S-shaped pattern (*25–27*). Notably, Sun and Ertz (*28*) use the system dynamics modelling framework to forecast ride-sourcing platforms' growth. They take into account various factors on growth, nonetheless, their model is macroscopic, which makes it impossible to capture the dynamics at the individual (microscopic) level.

    To fill this gap, we propose a novel microscopic co-evolutionary model featured by the fundamental growth factors to represent the day-to-day dynamics of the two-sided mobility market.

We implement an agent-based model to capture the interactions between the independent agents (rational utility maximizers) who get notified about ride-sourcing through marketing activities of platform, build their own experiences, share views with their social network (through word of mouth) and, accordingly, update their perceived utility at the end of each day. While the platform competes for travelers with alternative travel modes (like public transport) in a two-sided monopoly marketplace, on the other side, it competes for drivers with alternative occupations in the labor market.

The novelty of our approach lays in the S-shaped learning function, which we advocate to be the framework capable to reproduce how agents build their experience, adapt, and stabilize their behavior with time. Indeed, experiences and effects cumulate in the perceptions of the agents and are smoothed for a stable, yet sensitive behavior. Furthermore, their perceived utility is composed of: experienced, word of mouth and marketing components, which are adjusted separately. Notably, due to the nature of our model, agents are not fully optimal, and their decisions are often biased by opinions of peers (so-called word-of-mouth effect), marketing or individual taste variations (like willingness to pay). They need time to gradually build experience trying out the platform and learn its actual properties (e.g., waiting times or incomes) with unique individual behavioral trajectories. With more experience, their choices stabilize, but they remain sensitive to changes (e.g., increased waiting times, or lower incomes).

Our adjustment mechanism allows us to realistically model the variety of market entry strategies and observe platform growth at different stages. Furthermore, we are able to detect the positive and negative cross-side network effects in the platform lifecycle.

The paper is organized as follows in the next section. we introduce the S-shaped model and how it is implemented in the MaaSSim simulator. In the Results, section we detail the experiment illustrating the proposed model in 400-day simulation of two-sided mobility market in Amsterdam. Finally, we discuss the results and conclude in the last section.

## 2. Methodology

We model the two-sided mobility market with an agent-based microscopic model. The bottom-up agent-based approach is a powerful modelling framework to simulate independent decision-makers (agents) with different tastes and preferences, as well as to capture the potential interactions between them (*29*). To this end, we use our MaaSSim simulator and extend it with a co-evolutionary model to represent the day-to-day dynamics of the two-sided mobility market. We introduce two classes of agents, representing two sides of the system and a platform as an intermediate agent macthing the demand to the supply. A pool of travelers and drivers, who are not formerly notified about our ride-sourcing platform, gradually become aware of the ride-sourcing through the platform's marketing campaign. When an agent (driver/traveler) gets notified, she starts to learn and adapt her behavior through endogenous and exogenous factors.

**MaaSSim – two-sided urban mobility simulator**
MaaSSim is an open-source, *python* agent-based simulator, reproducing the dynamics of two-sided mobility platforms (like Uber and Lyft) in the context of urban transport networks (*30*) available on the public repository for reproducible experiments[1]. It models the behavior and interactions of two kind of agents: (i) travelers, requesting to travel from their origin to destination at a given time, and (ii) drivers supplying their travel needs by offering them rides. The interactions between the two types of agents are mediated by the platform, linking the demand to supply through a matching algorithm (we use the "*first-dispatch*" protocol for matching, which simply pairs the traveller with the closest idle driver, who is expected to have the shortest travel time (*31*)). Both supply and demand are microscopic. For supply, this pertains to the explicit representation of single vehicles and their movements in time and space (using a detailed road network graph), while for demand, this pertains to the exact trip request time and destinations defined at the graph node level. In MaaSSim our agents everyday make decisions to join the system (supply the demand as a drivers and/or travel to their destinations as the platform clients). They collect experiences on the travel times, waiting times and reliability (as travelers) or on the profitability of working for the platform (as the drivers). The model is non-deterministic, no only due to the inherent microscopic complexity but also due to probabilistic participation decisions of agents, thus daily realizations may change day-to-day significantly.

---
[1] https://github.com/RafalKucharskiPK/MaaSSim

**Platform**
The platform in our simulation executes a given market entry strategy, which, apart from pricing (per kilometer trip fare), may include discounts, marketing campaigns and commission rate modifications. While the platform everyday pursues the specific strategy, based on the above-mentioned tuning levers, it seeks to maximize its profit, which comes from the commission fees. In short, the platform aims to maximize the number of travelers on the demand side and serve this demand with a supply sufficient to provide a quality that leaves the travelers satisfied. We assume the platform operating costs to be fixed, except for the marketing campaign, which is paid per target agent per day.

**Traveler**
Each notified traveler *r* in our model on day *t* selects among two alternatives from a binary choice set $C_r=\{rs, pt\}$ including public transport (*pt*) and a new ride-sourcing mode of transport (*rs*). While the utility of public transport is fixed (and formulated with a typical access/egress, waiting times, transfers, etc.), the ride-sourcing utility is composed of multiple components, which are adjusted everyday as detailed below.

**Driver**
Similarly, each notified driver *d* makes a choice from a binary choice set $C_d=\{w, nw\}$ which includes working (*w*) and not working (*nw*) alternatives. By choosing not to work, a driver opts for the reservation wage which is the expected wage that she receives in the alternative labour market. The utility of working for the platform is adjusted every day in the same way as for the traveller.

**Agents' awareness**
An agent starts considering the new mode in her/his choice set only after being notified about it. We assume that this takes place during the marketing campaign. The marketing campaign gradually reaches to consecutive agents (depending on its intensity and duration), who include the new ride-sourcing mode in their choice sets. We control this process by determining the probability that each potential client is exposed to the marketing content everyday, which depends on the intensity of the marketing campaign and is reflected in its costs.

**Perceived utility components**
For any notified traveler and driver $i$, we propose the generic perceived utility ($U$) formulation composed of three components, namely: experienced utility ($U^E$), word of mouth utility ($U^{WOM}$) and marketing utility ($U^M$):

$$U_{i,t} = \beta_i^E \cdot U_{i,t-1}^E + \beta_i^M \cdot U_{i,t-1}^M + \beta_i^{WOM} \cdot U_{i,t-1}^{WOM} + ASC + \varepsilon_i \tag{1}$$

While all components are updated daily, the first one is endogenous and comes directly from the experience of agents obtained from the microscopic MaaSSim simulations. Drivers experience the actual income after each day of simulation, while travelers experience the travel times and costs of their trips. These experiences, along with other components of perceived utility, are diffused through the agents' social network through the word-of-mouth effect (*WOM)*. Finally, the marketing (apart from making agents aware of the new mode) generates a positive utility for the agents exposed to the platform's marketing campaign. These three components of perceived utility provide a rich representation of the actual dynamics driving the platform growth process. The β's in the formula reflect the relative weights of respective utility components (ensuring that $\beta_i^E, \beta_i^M, \beta_i^{WOM} > 0$ and $\beta_i^E + \beta_i^M + \beta_i^{WOM} = 1$), the alternative-specific constant (*ASC*) captures the effect of unobserved factors on the perceived utility of alternatives and $\varepsilon$ is the random utility error term. In such form, the utility is consistent with the discrete choice theory and can be applied e.g., in the logit model (like in this study).

**S-shaped learning and adaptation**
The key element of the proposed model lies in the following adjustment mechanism which allows us to realistically represent the agents' dynamics specific to the platform growth. Agents (drivers and travelers) update their perceived utility of ride-sourcing platform according to their past experiences, social influences and platform marketing activities. They learn and adapt their choices day-to-day. Here, instead of exponential smoothing (*23*, *24*), we follow Murre (*25*) and propose a more adequate formulation of the so-called S-shaped learning curve. Fig. 1 provides a basic idea of this model and its interpretation.

The adjustment process can be seen as moving along the S-shaped curve with each of the three components of the utility. The positive experience increases the utility and the negative experience decreases it. Triggered by the consecutive positive experiences, learning proceeds slowly for agents who already have sharp, extreme opinions, and is fast for neutral agents. Notably, learning can go both directions: an agent with no experience can learn the positive utility and eventually stabilize his expectations at high values. Yet later she can cumulate the opposite experiences (when the system performance deteriorates – as in the last stage of our experiment) and decrease expected utility gradually moving on the S-shaped curve to the left. Similarly, the decline will proceed slowly first and speed up only after a considerable number of negative experiences are cumulated. In this way, contrary to exponential smoothing, we can stabilize the agents' behavior and, at the same time, remain sensitive to system changes.

We cover two different phenomena with the S-shaped function: learning and effect, as distinguished by Haurand et al. (*26*, *27*). Our agents learn the exogenous system properties (profits for the drivers and travel times for travelers) and become affected by word of mouth and marketing. While new experiences trigger the learning, the word of mouth and marketing effects are provoked upon exposure. We assume word of mouth effect can be both positive and negative, depending on whether we share viewpoints with someone who has greater or lower utility than ours. For the marketing we consider only a positive effect.

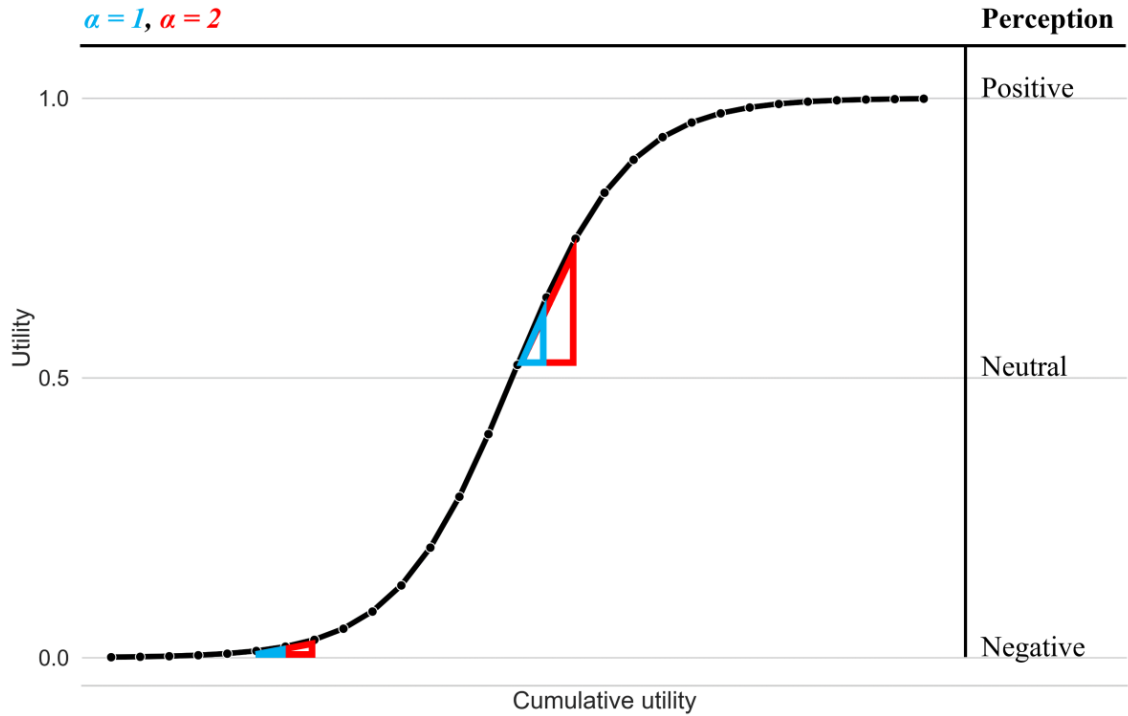

**FIGURE 1 S-shaped curves used to update utility perceptions in our experiments. The model with higher sensitivity (red) results in greater adjustment in comparison to the less sensitive model (blue). Notably, the sensitivity depends also on the position on the S-shaped curve, the updates are low for both highly negative and positive utilities and high when opinions are neutral.**

We formalize the above S-shaped model with the following formulations, which updates each component $c$ of utility as follows:

$$U_t^c = \frac{1}{1 + exp(\beta \cdot CU_t^c)} \quad (2)$$

$$CU_t^c = \ln\left(\frac{1}{U_{t-1}^c} - 1\right) + \alpha \cdot \Delta c_{t-1} \quad (3)$$

The utility ($U_t$) is every day updated through new experiences and exposures as a function of cumulative utility ($CU_t$) with a generic form of the sigmoid function (eq. 2). Parameter $\beta$ (the logistic growth rate) shapes the curve (and controls the sensitivity to the new experiences). The cumulative utility on day $t$ is computed with the (eq. 3), where we first obtain the cumulative utility of the previous day through the inverse sigmoid function and then add the utility difference ($\Delta c$) on day $t-1$ weighted with the sensitivity parameter $\alpha$, determining (along with $\beta$) how fast to move on the S-shaped curve. When there is no new experience or exposure, the utility difference will be equal to zero, which results in unchanged utility.

The above formulation is generic to represent various kinds of learning new experiences and being exposed to various effects. Below, we introduce the specific formulas for three components of utility adjusted in our model. For the endogenous part of the model, we adjust the drivers' and travelers' experienced utility as follows:

$$\Delta c_{d,t}^{E,w} = \frac{RW - I_{d,t-1}}{RW} \quad (4)$$

$$\Delta c_{d,t}^{E,rs} = \frac{U_{r,t-1}^{pt} - U_{r,t-1}^{rs}}{U_{r,t-1}^{pt}} \quad (5)$$

Experienced cumulative utility of drivers on day $t$ is updated through the relative difference between the reservation wage ($RW$) and the income experienced on the previous day ($I_{d,t-1}$). Similarly, travelers adjust their experienced cumulative utility on day $t$ through the relative difference between the utility of $rs$ and the alternative mode utility ($pt$). Note that both reservation wage and $pt$ utility are fixed, and the experiences can vary substantially day-to-day due to complex, non-deterministic interactions among agents in the microscopic simulations. Thanks to the S-shaped framework, such fluctuations both cumulate in the perceptions of the agents and are smoothed for a stable, yet sensitive behavior.

Analogously, we use the S-shaped curve to formulate the effects of word of mouth and marketing. For the word-of-mouth, we assume random pairwise interactions among the agents, who share their perceived utility with each other (bi-directionally). Expressed as $p_{i,j}^{WOM}$ the probability that agent $I$ shares her opinion with agent $j$ at day $t$ – which can be derived from density and intensity of underlying social networks and the links among agents. Here, we used synthetic uniform probability, which can be replaced with the actual structure of the social network. Influenced by the exchange of views with their peers, agents adjust their word-of-mouth component of utility:

$$\Delta c_{i,t}^{WOM} = p_{i,j}^{WOM} \cdot \left( U_{i,t-1}^{WOM} - U_{j,t-1} \right) \quad (6)$$

The marketing spreads uniformly to all the agents (target clients) and accumulates in time over the period of the marketing campaign. Before and after the campaign, the marketing effect is constant. We update the cumulative utility for marketing as follows:

$$\Delta c_{i,t}^{M} = p_{i}^{M} \cdot \left( U_{i,t-1}^{M} - 1 \right) \quad (7)$$

We assume that a marketing produces a positive effect on each exposure. Parameter $p_{i,j}^{M}$ is the probability that agent $i$ is exposed to marketing effect.

While experienced utility is learned, WOM and marketing utilities are obtained upon exposure as effects. Hence, we treat them differently. We assume that learning starts from the negative perception (from the lower tail of the curve) and with several experiences agents may either stabilize with a negative viewpoint or gradually move towards the positive one. Notably, this initially negative perception of the new mode, did not prevent agents in our study to become loyal clients with high perceived utilities. Such approach to simulate learning is substantiated by the literature (*25*, *32*). On the contrary, for the effects of WOM and marketing erect a strong initial effect which leads to positive or negative attitude at the end (*26*, *27*, *33*).

Accordingly, we assume at the beginning utility of WOM and marketing of the new mode are neutral, i.e. $U_0^{WOM}=0.5$ and $U_0^M=0.5$, while the initially null experience translates into $U_0^E=0$. Nonetheless, the adjustment process for any of the components can end in either a stable phase, where strongly negative or positive utility prevails, or a sensitive phase where utility is more unreliable (as illustrated in Figure 2).

**Choice probabilities**
For both drivers and travelers, we propose the logit model, which yields the choice probabilities as follows:

$$P_{r,t}^m = N_{r,t} \frac{exp(\mu \cdot U_{r,t}^m)}{\sum_{m \in C_r} exp(\mu \cdot U_{r,t}^m)} \qquad (8)$$

$N_{r,t}$ is a binary variable set to zero and switching to one as soon as an agent gets notified. Mind, that the above classical form of the logit model is also S-shaped, yet this is independent from the S-shape adaptations applied to the utility.

**Method summary**
We introduce a microscopic model capable to represent all the phenomena crucial to reproduce two-sided mobility platform rise and fall. Agents first become aware of the new mode and include it in their choice-sets. They have neutral opinions on the mode and have no experience with it. They receive opinions from their peers and through marketing channels and cumulate them in their utilities. They explore the new mode by using it and build their own experience. With time, they become more experienced and stable with their perception of the new mode, however, new feedback from any components of the utility can reverse the already stabilized perceptions. Negative opinion received via word-of-mouth, negative image brought by marketing campaign (e.g., the recent Uber scandal), or negative experience can reverse this trend. Like in our experiments, where drivers revert from platform due to reduced profits after platform has increased the commission fee. Despite positive marketing image, travelers experience lower utility (they need to wait longer since too few drivers work for the platform) and spread it with word-of-mouth to their peers, causing outflow of demand from the platform. All above modelled with a single, closed-form formulas capable of reproducing rich, multi-stage market evolutions.

3. **Results**

**Experimental design**
We run our experiments in Amsterdam, the Netherlands, a medium-size city (below 1 million inhabitants) with high-quality public transport. On the demand side, we consider a pool of 2000 travelers choosing between PT and ride-souring everyday to travel from their origin to destination. We sample these 2000 trips from the real-world Albatross tripset (*34*) filtering for trips longer than 2 km – which allows for an empirically plausible background. For each trip request, we query for the public transport alternative and obtain the detailed trip utility (composed of the classical access/egress times, waiting, transfers and fares) with OpenTripPlanner on a detailed OSM graph and GTFS timetable. Similarly, on the supply side, we assume a pool of 200 potential drivers who would choose between working as a platform driver and alternative occupations with reservation wage of 10.63[€/hour] (based on the minimum daily wage in the Netherlands (*35*)). The revenue of drivers working for the platform is calculated as the trip fare minus the platform commission fee. The operational costs (fuel, depreciation costs, etc.) of the drivers amount to 0.25 [€/km] which is deducted from revenues to obtain the drivers' profit.

The simulation time for each day is 4 hours, during which we simulate ride-sourcing services: travelers request the rides, and the platform matches them with the closest vehicle. The speed of ride-sourcing vehicle is set to the flat 36 [km/h]. We consider the trip fare of 1.2 [€/km] with a minimum fare of 2 [€] for the ride-souring, which is the actual average fare in Amsterdam (based on the Uber price estimator (*36*)). For the ride-sourcing utility, the waiting time weight is 1.5 times greater than the in-vehicle travel time (*37*)). We assume the trip requests do not change day-to-day (travelers have fixed origins, destinations and departure times) and drivers start their shifts from the same positions everyday (drawn randomly at the beginning of the simulation).

In our experiment, we fix the corresponding utility weights to $\beta_i^E = 0.80$ (for utility of experience), $\beta_i^{WOM} = 0.18$ (for utility of word of mouth) and $\beta_i^M = 0.02$ (for utility of marketing) for all agents. This is in-line with the findings on the actual platform-growth trajectories, fueled first with marketing, secondly with WOM, yet mainly with a positive experience in a cross-side network effects. We consider the diffusion speed of 10% for both marketing and WOM, i.e., probability of being exposed to marketing information during the campaign is 10%. Each agent has 10% probability to exchange views with some other agent every day. We do this by sampling 10% of agents every day and randomly assigning them in pairs among which they exchange views. The patience threshold of travellers to be matched is set to 10 minutes, after which they leave unsatisfied (if no driver is available to supply their demand). Such unfulfilled requests yield extra disutility for the platform (lost revenues) and for the travelers (negative experience).

**Platform strategy**

The platform seeks to maximize its profit. We propose a market entry strategy consisting of six consecutive stages over the 400 days of simulation (summarized in Table 1). Platform starts with 10% commission at the *kick-off stage*, which is followed by a 40% discount on trip fares in the *discount stage*. The discounting scheme is a specific one: the platform reduces the trip fare only for the travelers, not for the drivers (who receive the full fare minus commission). Moreover, we offer the discount only for those travelers who are not yet loyal to the platform (i.e. their probability to use ride-hailing is below 50%). Thus the discounting schema is a subsidizing by the platform, without negative impact on the drivers (contrary to commission fee) and positive impact on the travelers.

Nonetheless, the platform is not successful in user acquisition, until both supply and demand sides are not notified yet about the new mode of transport ($N_{r,t}$ in eq. 8 is zero). This changes in the *launch stage,* the 50-day marketing campaign, to launch the platform, triggers the user acquisition through not only notifying them but also creating a positive image of the platform (via eq. 7). We assume that marketing activity costs 5 [€/agent/day], which adds up to 1100 [€/day], considering the total size of supply and demand (2200 agents). It should be noted, each day, the campaign only reaches a random 10% of target agents. After 50 days of marketing, when all agents are already notified, the platform quits the campaign upon reaching a fast growth pace in the *growth stage*. When it reached sufficient size, the platform decides to end discounting on the day 200 which leads to the *maturity stage*, when it stabilizes profits, market shares and agent behavior. In the last 100 days of the proposed scenario, the platform opts for a greedy move via increasing the commission rate to 50% to maximize the profit during the *greed stage,* which has catastrophic consequences, as detailed below.

**TABLE 1 The rise and fall of the two-sided mobility platform in Amsterdam, expressed with the six-stage market entry strategy adopted by the platform in our experiments.**

| Day | Stage number | Name | Marketing | Commission | Discount |
|---|---|---|---|---|---|
| 0 – 25 | I | Kick-off stage | – | 10% | – |
| 25 – 50 | II | Discount stage | – | 10% | 40% |
| 50 – 100 | III | Launch stage | 5 [€/agent/day] | 10% | 40% |
| 100 – 200 | IV | Growth stage | – | 10% | 40% |
| 200 – 300 | V | Maturity stage | – | 10% | – |
| 300 – 400 | VI | Greed stage | – | 50% | – |

We apply our simulation on the Amsterdam road network from OpenStreetMap. Figure 1 illustrates how agents become first notified and then active members of the platform in the four consecutive snapshots of the simulations.

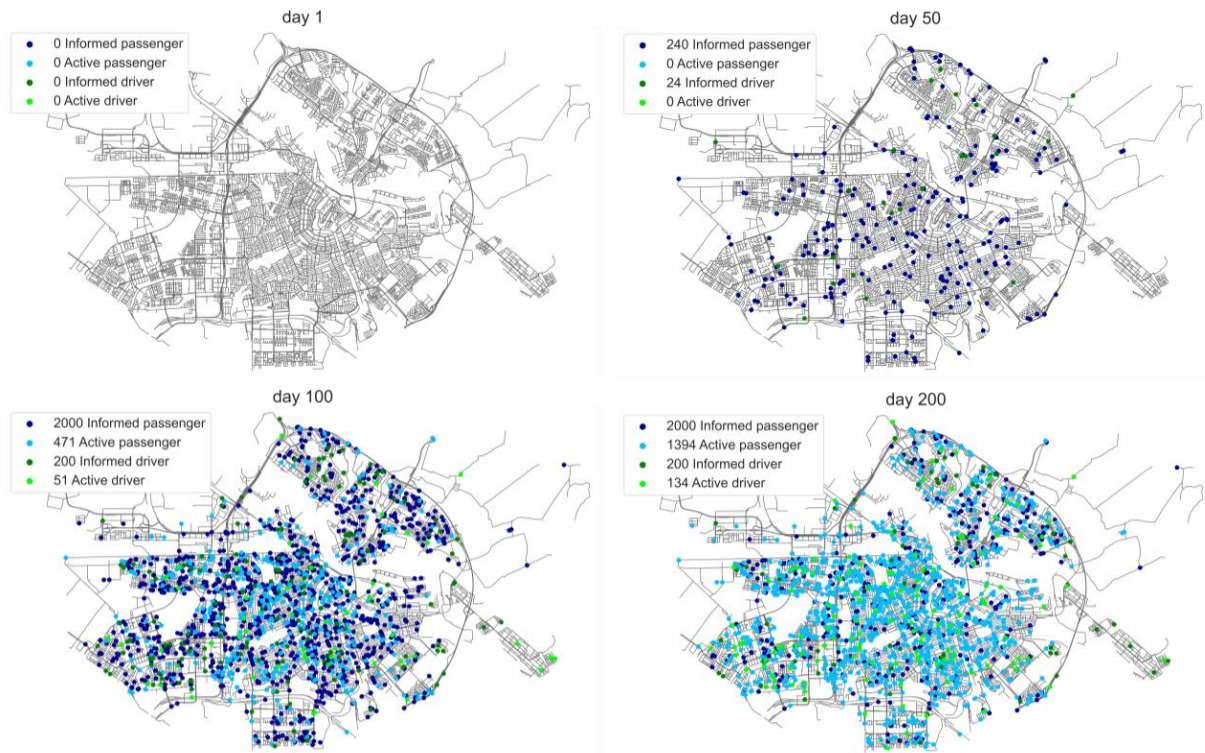

**FIGURE 1 Detailed road network used in experiments (Amsterdam, The Netherlands) snapshotted at four selected days of the simulation. Dots denote drivers (green) and travelers (blue), marked with dark color when they are notified and with light color when they participate in the platform.**

**Findings**

While the strategies outlined in Table 1 are implemented on the whole supply-demand scale, each agent follows a distinct, unique evolution path depending on her exposition to the marketing, word of mouth received from her peers and, most notably, her own experiences gained while participating in the platform. Figure 2 provides a comprehensive illustration on participation probability of agents through the perceived utility components over 400 days. Indeed, the utility coming from the direct experience is playing a significant role. Yet, it is the marketing that initially incentivizes the agents to try out the new mode. Word of mouth, however, for each agent represents the general (social) network view on the ride-souring which can considerably differ among agents.

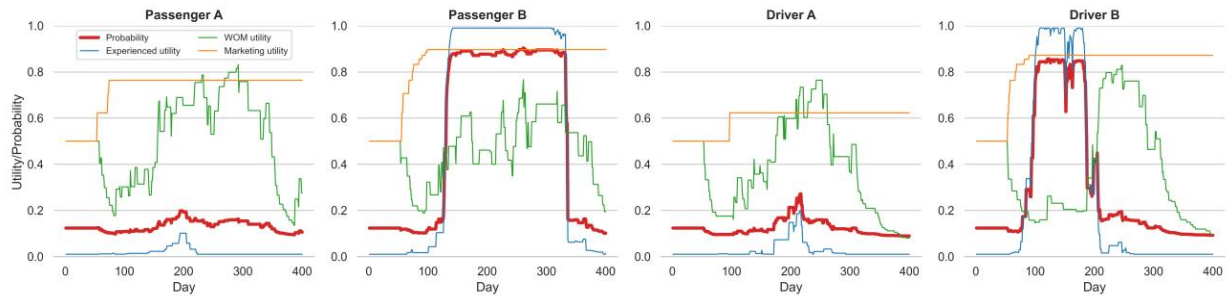

**FIGURE 2 Temporal evolution of choice probability (red) and three components utility for four sampled agents with various temporal evolutions. Passenger A is initially exposed to positive marketing (orange) and negative WOM (green) effects. However, positive WOM pushes her to try out the platform, she does not cumulate positive experiences (blue), and her probability of using platform remains low. Passenger B is more exposed to marketing compared to passenger A. She accumulates considerable positive experiences with the platform, and her participation probability stabilizes around 0.9, which drops to 0.1 at greed stage. Driver A goes through the a similar evolution path with to the passenger A. In which she is exposed to marketing only once. Driver B reaches a high probability of using platform despite negative WOM effects. However, she opts out from the system already in its maturity stage presumably due to bad experiences.**

To see how these individual experiences and trajectories of the agents' evolution translate into the rise and fall of the platform, Figure 3 provides system-wide averages. For the six stages of market entry strategy, it illustrates the three components of perceived utility (system-wide means) and the resulting market share. Since demand and supply are naturally balancing each other, thanks to our model structure, they produce very similar trends. Agents with no experience and neutral attitude toward word of mouth and marketing activities of platform, remain inattentive to the new mode until getting notified (stage III). By adding the ride-souring to their mode choice set, they start to occasionally explore it and build their experience. Later (stage IV), the value creation through the positive cross-side network effect, fueled by discounts, speeds up the platform growth. This results in many agents who preferably opt for ride-souring and find the new mode valuable. When platform ends the discounts (stage V) it reaches stability with 60% of market share. Intending to maximize its profit in short while, the platform dramatically raises the commission rate (stage VI), which turns to a tragedy losing a great market share on both demand and supply side.

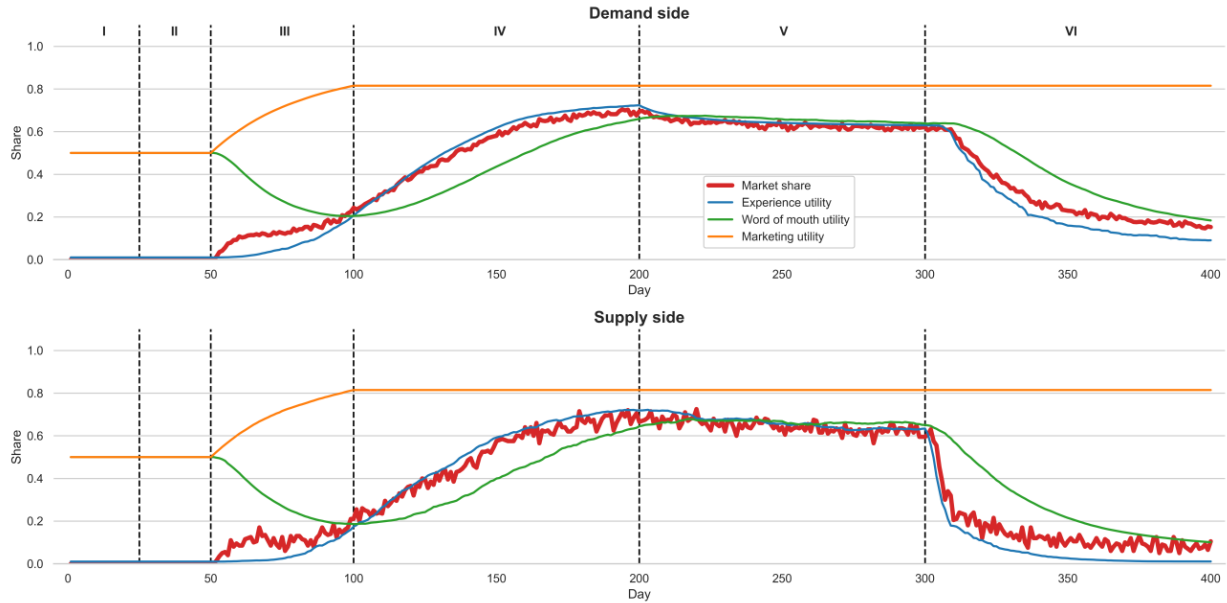

**FIGURE 3 Six stages and 400 days of the evolution of the market entry strategy in Amsterdam. The market share (red) results from individual agents' decisions, based on utility, composed of three components: marketing (orange), word of mouth (green) and experience (blue). In the first two stages, despite offering discounts from 25$^{th}$ day, the platform fails to launch. Marketing campaign (III) launched on day 50 attracts both supply and demand, which allows the platform to reach 20% of market share before the campaign ends on day 100. Nonetheless, the strong word of mouth effect sustains the steady growth (IV) until the stabilization before the 200$^{th}$ day. When stabilized, the platform stops offering discounts (V), which only slightly affects its stable market shares. The greed (VI), however, leads to a significant drop in market share from day 300, when the platform decides to collect 50% from the drivers as the commission fee.**

We investigate the performance of the platform for the six stages of the market entry strategy through the key performance indicators (KPIs). On the Figure 4, we see the average waiting time for travelers composed of the average pick-up time and the matching time. While in the early stages the waiting time is driven mostly by pick-up time, significantly increased matching time plays a substantial role at the greed stage. Illustrated on the right side of Figure 4, the profit drops below the reservation wage from day 300 for drivers. This happens initially due to high rate of commissions and gets worse with increased operation costs. The average idle time of drivers reaches its maximum in the stage V because of shortened pick-up times. For the same reason, travelers are experiencing minimum waiting times at the same stage.

We depict passenger and vehicle kilometers in Figure 5 (a) through scatters and bold smoothed curves. These curves follow the market share trends. While, the curve for vehicle kilometer move slightly above passenger kilometer curve, both of them look dispersed and unstable at the last stage. Furthermore, passenger kilometer curve jumps unexpectedly on the first few days of stage VI because with the commission rate raise from day 300 drivers rapidly start to opt out of the platform due to negative experiences. Yet, the negative wave meets the travelers, few days later, only when the platform loses considerable market share on the supply side. This brings with itself higher waiting and matching times.

Figure 5 (b) throws light on platform loss and gains in different stages. The revenue is below zero up to day 200, since the money spent on discounts is more than the total gain from commission fees. The platform breaks even and compensates the costs of marketing campaign after 250 days of operation. From beginning of the greedy stage (day 300), profitability increases due to raised commission rate and yet then declines, because of reduced market share.

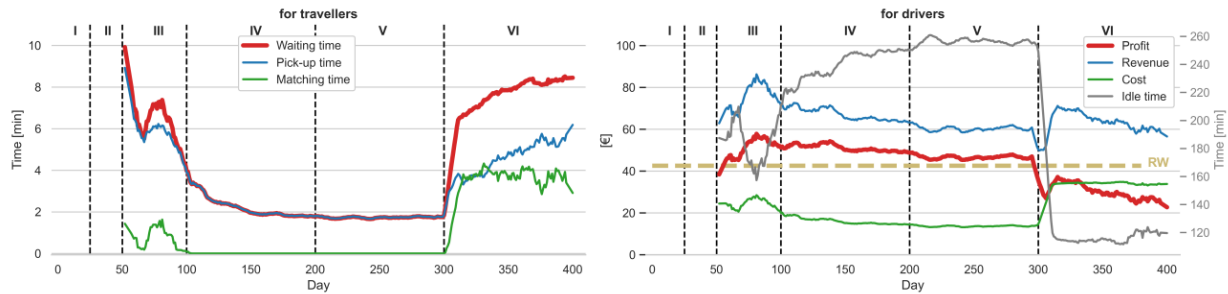

**FIGURE 4 Evolution of key performance indicators for the supply (left) and demand (right). Travellers' waiting times (left) steadily drop to two minutes per traveler after half a year and quickly increase again in the greed stage. In the early stage the pick-up is the main component of waiting, while in the latter the matching time becomes equally significant. The drivers' profits remain (red on the right) above the reservation wage – despite continuous growth – until it drops when platform decides to collect 50% commission fee.**

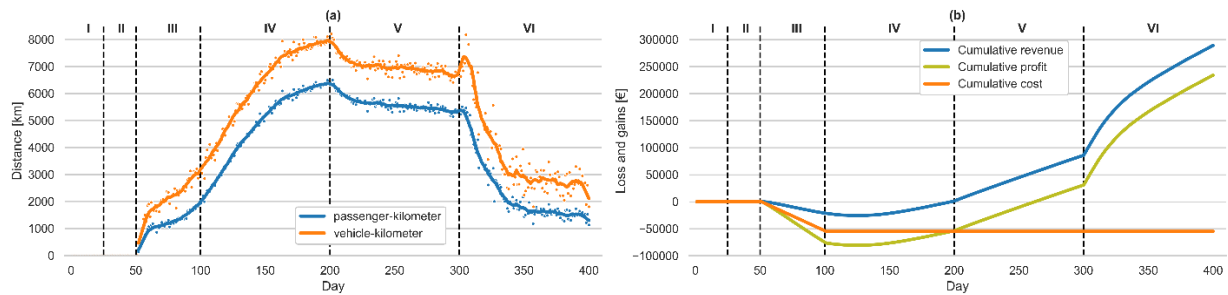

**FIGURE 5 (a) Passenger (blue) and vehicle (orange) kilometers' during the six stages of evolution. (b) Cumulated costs of marketing (red) and revenues (green). The operations become profitable after 250. The increased commission after 300 days initially increases the profits, yet quickly the daily revenues become lower than in the maturity stage.**

## 4. Discussion and conclusions

In this research, we propose a novel approach to model the market entry strategy of two-sided mobility platforms. Using MaaSSim agent-based simulator, we develop a co-evolutionary model to represent the day-to-day dynamics of the two-sided mobility market with agents of both supply and demand sides. Agents are rational decision-makers maximizing their perceived utility. We introduce the total utility as a composition of: experience, word of mouth and marketing components. We run a 400-day simulation with a predefined six-stage market entry strategy of the ride-sourcing platform. Agents, operating on a road-network of Amsterdam, initially, have neutral opinions on the new ride-sourcing mode and have no experience with it. As they get notified, they include the new mode in their choice-set and start to build their own experience and share view with their peers. To steer those supply-demand interactions toward the profitability, the platform controls: trip fare, commission rate and discounts, and launches marketing campaigns.

The proposed model enabled us to simulate a strong positive cross-side network effect at the microscopic level, which is crucial for platform growth. At the beginning of the simulation, there are few agents on both sides of the platform. The service performance is thus low, with long pick-up distances and matching times. Hence, travelers experience long waiting times and there is a low number of trips to be supplied by the drivers. However, as some travelers and drivers decide to remain in the platform (due to positive marketing campaign), the system starts the value creation which feeds the positive cross-side network effect. Indeed, in a feedback loop, the travelers' waiting time decrease and more travelers opt for the platform, which results in more trips incentivizing drivers to join the platform. While the value creation is small in the early stages, it becomes significant as the platform reaches considerable market share.

In the maturity stage, when the marketing campaign is finished and discounts are no longer offered, the market share of platform decreases slightly. Even though, the value created for the platform through network effect does not disappear. This should be the main reason behind the large subsidies that transportation network companies apply at the early adaptation phase. By reaching a high market share, they hope to keep their users satisfied on both sides and move toward the profitability. However, platforms can also collapse as a result of inappropriate strategies. For instance, in the greed stage of our market entry strategy, the platform raises the commission rate to 50% to supplement the profit. This activates the negative cross-side network effect, and platform loses a large amount of market share in short time.

Furthermore, we found that each agent follows a distinct, unique evolution path depending on its exposition to the marketing, word-of-mouth received from the peers and, most notably, own experiences gained while participating in the platform. A traveler or driver may decide not to participate in the platform, against her benefits, only because of negative views being exchanged in her social network. The opposite is also probable. A traveler or driver may decide to stay in the platform, just because of the positive opinions around her, despite negative experiences. Moreover, heterogeneous preferences could be set for some agents through utility weights, which would allow one to observe adaptation process for specific preference-genes.

We believe that the proposed modelling framework offers a richer and more solid representation of the system. Here, we illustrate its capabilities with a comprehensive case study. In the future, it can be applied to a variety of open research problems, offering a new insights and an experimental underpinning. This refers to, e.g., minimal wages, platform competition, congestion charging, driver number caps (like those introduced in NYC), ride-pooling or (notable in the light of recent revealed details) lobbying towards ride-sourcing platforms at the top political levels.

**ACKNOWLEDGMENTS**
This research is funded by National Science Centre in Poland program OPUS 19 (Grant Number 2020/37/B/HS4/01847) and by Jagiellonian University under the program Excellence Initiative: Research University (IDUB).

**AUTHOR CONTRIBUTIONS**
The authors confirm contribution to the paper as follows: study conception and design: F.G., R.K., ; data collection: F.G.; analysis and interpretation of results: F.G., R.K.; draft manuscript preparation: F.G., R.K. All authors reviewed the results and approved the final version of the manuscript.